\documentclass[11pt]{article}
\usepackage{graphicx}
\usepackage{hyperref,overpic}

\usepackage[margin=1in, paperwidth=8.5in, paperheight=11in]{geometry}

\usepackage{color}
\definecolor{red}{rgb}{0.8500, 0.1250, 0.0480} 
\definecolor{blue2}{rgb}{0, 0.4470, 0.7410}
\definecolor{green}{rgb}{0.4660, 0.6740, 0.1880}
\definecolor{gray}{rgb}{0.7, 0.7, 0.7}

\hypersetup{
	colorlinks=true,
	linktoc=all,
	linkcolor=blue,
	citecolor=blue2,
}

\usepackage{fancyvrb,xcolor}

\usepackage[font=sf,labelfont={sf,bf},margin=4mm]{caption}

\usepackage{natbib}
\usepackage{sectsty}
    \chapterfont{\sffamily\huge}
    \sectionfont{\sffamily\large}
    \subsectionfont{\sffamily\normalsize}
    \subsubsectionfont{\sffamily\normalsize}

\let \oldsection \section
\renewcommand{\section}{\vspace{-5pt plus -5pt}\oldsection}
\let \oldsubsection \subsection
\renewcommand{\subsection}{\vspace{-5pt plus -5pt}\oldsubsection}

\begin{document}

\title{\bf{3D Printing of Fluid Flow Structures}\footnote{Part of this work was supported by the US Air Force Office of Scientific Research (Program manager: Douglas Smith, Grant number FA9550-13-1-0091).}}

\author{{\bf Kunihiko Taira} and {\bf Yiyang Sun} \\
Department of Mechanical Engineering\\
Florida State University \\
ktaira@fsu.edu and ys12d@my.fsu.edu\\
~\\
{\bf Daniel Canuto}\\
Department of Mechanical and Aerospace Engineering\\
University of California, Los Angeles\\
dannycanuto@g.ucla.edu}

\date{\small Updated: July 05, 2017}

\maketitle

\begin{abstract}
We discuss the use of 3D printing to physically visualize (materialize) fluid flow structures.  Such 3D models can serve as a refreshing hands-on means to gain deeper physical insights into the formation of complex coherent structures in fluid flows.  In this short paper, we present a general procedure for taking 3D flow field data and producing a file format that can be supplied to a 3D printer, with two examples of 3D printed flow structures.  A sample code to perform this process is also provided.   3D printed flow structures can not only deepen our understanding of fluid flows but also allow us to showcase our research findings to be held up-close in educational and outreach settings.\\
  

\end{abstract}

\section{Introduction}
\label{sec:intro}

For students starting their training in fluid mechanics, one of the major challenges they encounter is that most fluid flows are not readily visible.  This is precisely why the black-and-white, 1960s collection of National Committee for Fluid Mechanics Films that contains beautiful flow visualizations to explain fluid mechanics, has been long cherished as a great educational resource even to this date \citep{national1972illustrated}.  Over the course of advancement in fluid mechanics, the development of flow visualization techniques has been critical to enable experimental and computational investigations \citep{Smits12,Samimy03}.  The beauty of the visualized complex and delicate flow structures has attracted researchers to uncover their formation mechanisms and their effects on the overall dynamics.

In experiments, there are a number of techniques to visualize the flow field.  Since most fluid flows are transparent, fluid mechanicians have relied on visible media or tracers to highlight flow features.  Dye and bubbles have been extensively used to identify flow patterns \citep{Smits12}.  Presently, one of the de facto standards in capturing the velocity field is particle image velocimetry (PIV), which uses tracer particles and cross correlation techniques to determine displacement vectors \citep{Raffel07,Adrian10}.  While PIV was originally limited to two dimensions, development of stereoscopic and tomographic PIV techniques has enabled the extraction of full 3D velocity vectors along a plane and over a volume, respectively \citep{Prasad:EF00,Elsinga:EF06}.

Moreover, computational fluid dynamics simulations have allowed full access to a variety of flow field data and enabled visualization of coherent structures in flows, even down to the finest details.  The use of isosurfaces and structural identification schemes have provided us with deep insights into the formation of such structures over space and time \citep{Cummins15,Kajishima17}.  Vortex identification techniques, such as the $Q$ and $\lambda_2$ criteria, have revealed the formation and evolution of vortical structures \citep{Jeong:JFM95,Jiang04}.  Such identification techniques are also used for experimental data.
 
Here, we discuss the use of {\emph{3D printing}} to materialize flow structures based on computational and experimental flow field data.  3D printing \citep{lipson2013fabricated} has become widely available in industry and academia, enabling rapid prototyping of parts out of plastics, metals, ceramics, and other materials.  There are also emerging efforts to print biological organs and prosthetics \citep{Schubert159}.  In fluid mechanics, we can take advantage of these 3D printing technologies\footnote{Another major use of 3D printing is the fabrication of scaled models for wind and water tunnel testings.} to present representative flow structures to an audience, akin to material scientists or roboticists showcasing their sample solid structures or robots, respectively.  
  
This new technology offers an opportunity for fluid mechanicians to materialize coherent structures and share them with students in classrooms, technical conferences, public outreach activities, and even as artistic objects.  The ability to physically hold, touch, and view printed flow structures provides unique opportunities for firsthand studies of the flow structures in research and educational settings.

\section{Preparation of Data for 3D Printing}

We can print 3D models of fluid flow structures from computational or experimental flow field data.  One of the most straightforward approaches is to consider a 3D isosurface for printing.  Once the 3D structure is generated, its surface data can be exported to a CAD data format.  We provide a sample MATLAB script (see Code 1) that takes 3D flow field data and outputs a STereoLithography (STL) file, which is a file format widely used in 3D printing.  When this STL file is created, it can be sent to a 3D printer for fabrication of the 3D model.

Depending on the 3D printer specifications, care must be taken to check the fine-scale details of the model to be produced.  At the moment, sharp edges and small-scale flow features can be a challenge for 3D printing, especially those seen in turbulent flows. The material properties of the print medium influence models' maximum fineness.  For flows with multi-component structures, we can print each piece separately.  In the following section, we present two examples in which the 3D models are comprised of single as well as multiple pieces.   We note in passing that some printers can print in multiple colors, enabling color map projection onto a 3D model (e.g., pressure distribution).

\begin{table}[h!]
{\small
~\\
\hrule
\begin{Verbatim}[commandchars=\~\{\}]
~typeC function data2stl(x,y,z,g,isoc_val,fname)
   ~typeA% Authors: K. Taira, Y. Sun (FSU), D. Canuto (UCLA), Nov 14, 2016
   ~typeA% === inputs === 
   ~typeA%   (x,y,z) = spatial grid points (e.g., meshgrid format)
   ~typeA%   g = variable of interest defined over (x,y,z)
   ~typeA%   isoc_val = isocontour value used to create structure
   ~typeA%   fname = output STL file name (.stl will be appended)
   ~typeA% === output ===
   ~typeA%   fname.stl = STL format file (for 3D printing)

   ~typeA% output faces (f) and vertices (v)
   [f,v] = isosurface(x,y,z,g,isoc_val); 
   ~typeA% write to STL file; need stlwrite.m from Matlab file exchange
   stlwrite([fname,'.stl'],f,v)
~typeC end function 
\end{Verbatim}
\hrule
\vspace{2mm}
}
\caption*{\small Code 1: Example MATLAB code to export isosurface in STL format for 3D printing.}
\vspace{-3mm}
\end{table}

\section{Examples of 3D Printed Flow Structures}
\label{sec:example}

\subsection{Vortices behind a pitching low-aspect-ratio wing}

The formation of three-dimensional vortices behind a low-aspect-ratio rectangular wing is known to be complex \citep{Taira:JFM09}.  As the vortices develop from the leading-edge, trailing-edge, and tips of the wing, they interact while convecting downstream.  Unsteady wing maneuvers such as pitching and acceleration can also influence the vortex dynamics \citep{Jantzen:PF14,Jantzen:AIAA13}.   

\begin{figure}
\begin{center}
  \includegraphics[width=0.79\textwidth]{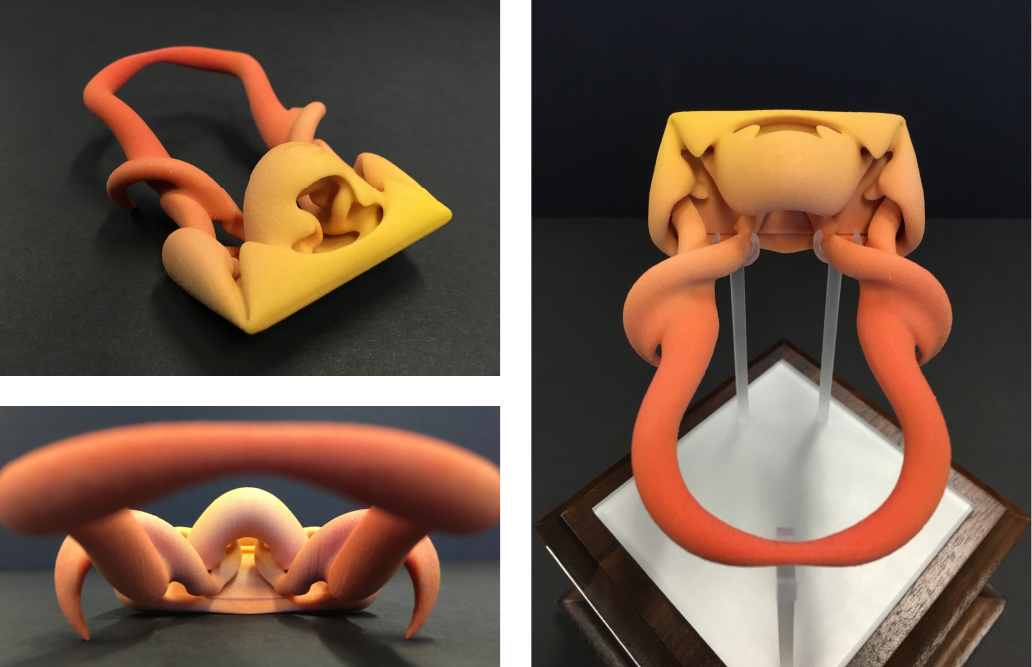} 
\caption{3D printing of vortical structures forming behind a pitching plate of aspect ratio 2 at $Re = 500$ \citep{Jantzen:PF14}.}
\label{fig:3dprint1} 
\end{center}     
\end{figure}

Here, we consider printing the data obtained from DNS of an unsteady incompressible wake behind a rectangular flat-plate wing of aspect ratio 2, undergoing a pitching maneuver at $Re = 500$ \citep{Jantzen:PF14}. The simulation is performed with the immersed boundary projection method \citep{Taira:JCP07,Kajishima17}, and the vortical structures in the wake are captured by a $Q$-criterion isosurface \citep{Hunt:CTR88}.  These structures and the flat-plate wing are printed with a 3D printer, as shown in Figure \ref{fig:3dprint1}.  The color projected on the model represents the streamwise coordinate to highlight the wake structure location with respect to the wing.

The 3D model reveals the intricate details of the formation of the leading-edge and tip vortices at an early stage of the dynamics.  Viewing the printed model from the rear, we can observe how the legs of wake vortices are intricately wrapped around each other while pinned to the top surface of the wing, satisfying Helmholtz's second theorem.  The ability to hold and examine the model provides firsthand insights into the complex vortex dynamics caused by the unsteady wing motion and the separated flow.

\subsection{Global stability mode inside a cavity}

We can also print structures from modal decomposition and stability analyses \citep{Taira:AIAAJXX,Theofilis:ARFM11}.  These flow structures hold importance in understanding the dynamics and stability of fluid flows.  In particular, the dominant global stability mode highlights how a perturbation in a flow can grow or decay about its base state.  

\begin{figure}
\begin{center}
  \includegraphics[width=0.79\textwidth]{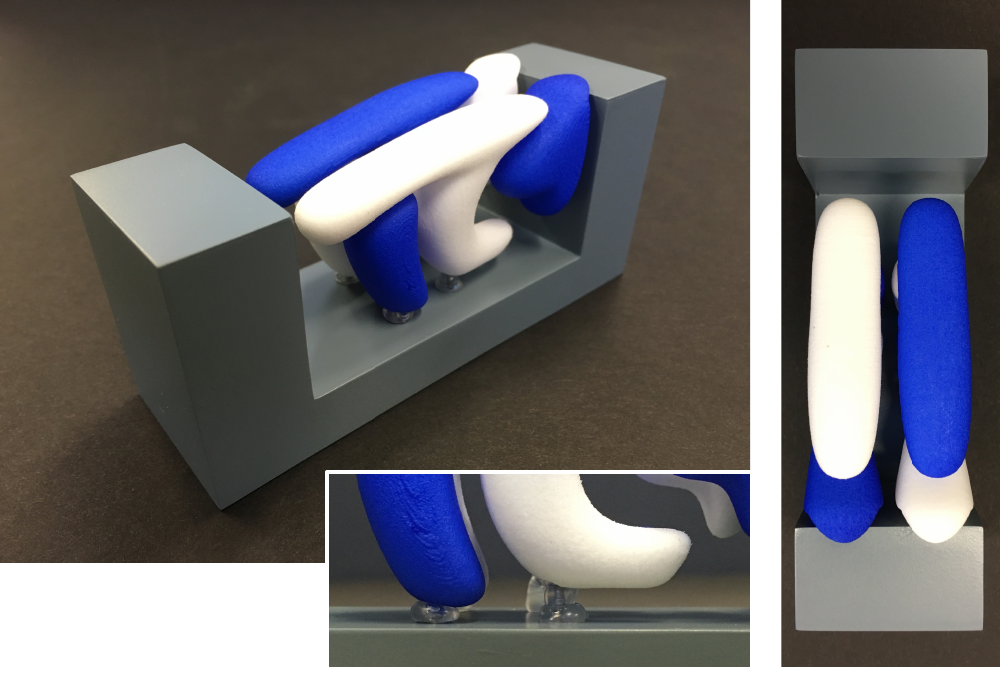} 
\caption{3D printing of the dominant biglobal mode for $M_\infty = 0.6$ and $Re = 1500$ in a rectangular cavity of aspect ratio $L/D=2$ \citep{Sun:JFMXX}.}
\label{fig:3dprint2} 
\end{center}     
\end{figure}

As an example, we 3D print the dominant stability mode determined from global stability analysis of compressible flow over a spanwise-periodic rectangular cavity at $Re = 1500$ and $M_\infty = 0.6$ with aspect ratio $L/D=2$ \citep{Sun:JFMXX}.  The stability analysis is performed here with a global stability analysis code (large-scale eigenvalue problem) developed upon the finite-volume solver CharLES \citep{Sun:JFMXX,Bres:AIAAJ17}.  Shown in Figure \ref{fig:3dprint2} are two sets of isosurfaces (positive and negative) of the dominant global stability mode in terms of the spanwise velocity with two different colors.  

We print these structures as separate pieces, which are not in contact with the cavity (fabricated separately).  Hence, thin metal wires and adhesives are used to hold the modal structures in place, as shown in the inserted figure.  The spanwise size of these structures indicates the wavelength of this particular mode, and the temporal frequency influences the size of the individual structures.  The structure of the mode can be studied to assess where sensors and actuators may be placed for flow control.  The 3D printed stability mode offers a chance to examine its structure up-close, which is not ordinarily visible in cavity flows.  The model can also facilitate the conveyance of this specialized concept of global stability to a general audience through physical contact and viewing.

\section{Concluding Remarks}
\label{sec:summary}

We have discussed the use of 3D printing to physically visualize fluid flow structures.  By having a materialized structure, we believe that such printed models can further deepen our understanding of fluid flows, and allow us to showcase our research in educational or outreach settings.  At the moment, most 3D printers cannot output models that have very fine structures, which may limit the Reynolds number of printable flows.  However, with continuous improvement of 3D printing technology, this limitation may be relaxed or eliminated in the future.  Moreover, use of transparent printing media and future reduction in cost may allow for complex flows resulting from turbulence to be physically printed.  We hope this short paper stimulates readers to try 3D printing of fluid flow structures obtained from their computational and experimental studies and share their models in various opportunities to highlight the beauties of fluid flows.

\bibliography{CFD_refs}  

\end{document}